\documentclass[12pt]{article}
\pdfoutput=1
\usepackage{amsmath,amssymb}
\usepackage{amsthm}
\usepackage{mathrsfs} 
\usepackage{graphicx}

\newcommand{\be}{\begin{equation}}
\newcommand{\ee}{\end{equation}}
\newcommand{\bea}{\begin{eqnarray}}
\newcommand{\eea}{\end{eqnarray}}

\newcommand{\gapp}{\mathrel{\raise.3ex\hbox{$>$}\mkern-14mugo
              \lower0.6ex\hbox{$\sim$}}}
\newcommand{\lapp}{\mathrel{\raise.3ex\hbox{$<$}\mkern-14mu
              \lower0.6ex\hbox{$\sim$}}}

\newcommand\lsim{\lesssim}

\newcommand\gsim{\gtrsim}

\newcommand\vev[1]{{\langle {#1} \rangle}}
\renewcommand\({\left(}
\renewcommand\){\right)}

\newcommand\eq[1]{Eq.~(\ref{#1})}
\newcommand\eqs[2]{Eqs.~(\ref{#1}) and (\ref{#2})}
\newcommand\eqss[3]{Eqs.~(\ref{#1}), (\ref{#2}), and (\ref{#3})}

\newcommand\mpl{M_{\rm P}}

\newcommand\dbibitem[1]{\bibitem{#1}}
\newcommand{\dlabel}[1]{\label{#1}}

\def\calp{{\cal P}}
\def\calr{{\cal R}}

\def\calpz{\calp_\zeta}

\newcommand\bfe{{\mathbf e}}

\newcommand\bfk{{\mathbf k}}

\newcommand\bfn{{\mathbf n}}

\newcommand\bfp{{\mathbf p}}

\newcommand\bfx{{\mathbf x}}

\newcommand\GeV{\,\mbox{GeV}}

\newcommand\Mpc{\,\mbox{Mpc}}

\newcommand\sub[1]{_{\rm #1}}
\newcommand\su[1]{^{\rm #1}}

\newcommand\mone{^{-1}}

\newcommand\mfive{^{-5}}

\newcommand\half{^{1/2}}

\newcommand{\fnl}{f\sub{NL}}

\newcommand{\dchil}{\delta\chi\sub L}
\newcommand{\dchis}{\delta\chi\sub S}
\newcommand{\calpzl}{\calp_{\zeta\sub L}}

\newcommand{\xls}{x\sub{ls}}
\newcommand{\zetal}{\zeta\sub L}
\newcommand{\zetas}{\zeta\sub S}
\newcommand{\kl}{{k\sub L}}
\newcommand{\ah}{{a_0H_0}}
\newcommand{\cgz}{C_2\su{GZ}}

\newcommand{\az}{A_\zeta}
\newcommand{\cz}{C_\zeta}

\newcommand{\ns}{n\sub s}

\begin{document}

\title{Generating  $\fnl$ at $\ell\lsim 60$}
\author{David H.\ Lyth\\Consortium for Fundamental Physics,\\ Cosmology and
Astroparticle Group, Department of Physics,\\ Lancaster University,
Lancaster LA1 4YB, UK}
\maketitle
\begin{abstract}  The CMB anisotropy at $\ell\lsim 60$ seems to have some special features which include (i) a dipole modulation and (ii)
a decrease in power. It is known that both of these
 effects can be generated if
 a curvaton-type field has a super-horizon perturbation.
 It is also known that this will  generate non-gaussianity $\fnl$ in the same range of $\ell$,  whose magnitude
has a lower bound coming from the magnitude of the observed CMB quadrupole. I revisit that bound in the present
paper, and point out that it may
  or may not be compatible with current data which should therefore be re-analysed.
\end{abstract}

\section{Introduction}

For multipoles $\ell \lsim 60$ (corresponding to large scales
$k\mone =\xls/\ell \gsim 200\Mpc$ where  $\xls=14,000\Mpc$
is the distance of the
last scattering surface) the CMB anisotropy seems to have some special features. One of these is a dipole
asymmetry \cite{wmap,planckisotropy}
\be
\Delta T = \(1+A(\hat\bfp\cdot \hat\bfn) \)\Delta T \sub{iso}(\hat\bfn)
, \ee
with $A=0.07\pm 0.02$. The effect seems to be a real one \cite{akrami}, despite an earlier suggestion to the contrary \cite{bennett}. It must be scale-dependent though, because observation requires \cite{notari,fh}  $A<0.0045$ (95\%
confidence level) for $\ell=601$-2048 corresponding to $k\mone \sim 10\Mpc$.
Another feature \cite{wmap,planckisotropy,planckcos}, shown in Figure \ref{cmb}, is a suppression in the
magnitude of the multipoles, below the level that would correspond to the $\Lambda$CDM model with constant
spectral index $\ns=0.96$ that provides a good fit at higher $\ell$.

Making the usual assumption that the CMB anisotropy is generated by the
 primordial curvature perturbation
 $\zeta$, these features correspond to
 \be
 \zeta_\bfk(\bfx)
= \( 1-\cz(k)+ \az(k) \hat\bfp \cdot \bfx/\xls  \) \zeta_\bfk
, \dlabel{approx} \ee
 and
\be
\calpz(k,\bfx) \simeq \( 1-2\cz(k)+ 2\az(k) \hat\bfp \cdot \bfx/\xls  \) \calpz(k)
\dlabel{calpofx} .\ee
This expression, and its meaning was given in \cite{p13}. The quantity
 $\zeta_\bfk$ is a  statistically homogeneous quantity, which fits the higher multipoles and has the constant
spectral index $\ns=0.96$ that is required by observation.
The quantity $\zeta_\bfk(\bfx)$ is also statistically homogeneous, but it is defined only within a  box at the position $\bfx$, whose size is much smaller than $\xls$. For that to make sense, I consider only
$k\mone \gg \xls$.
To account for the dipole asymmetry we need $\az(k)\sim 0.07$ for $k\mone\gsim \xls/60$,
but the  distribution of distant quasars \cite{hirata} requires
$\az<0.014$ (99\% posterior probability)
 on the  scale $k\mone\sim 1\Mpc$.

\begin{figure}
\includegraphics[width=\textwidth]{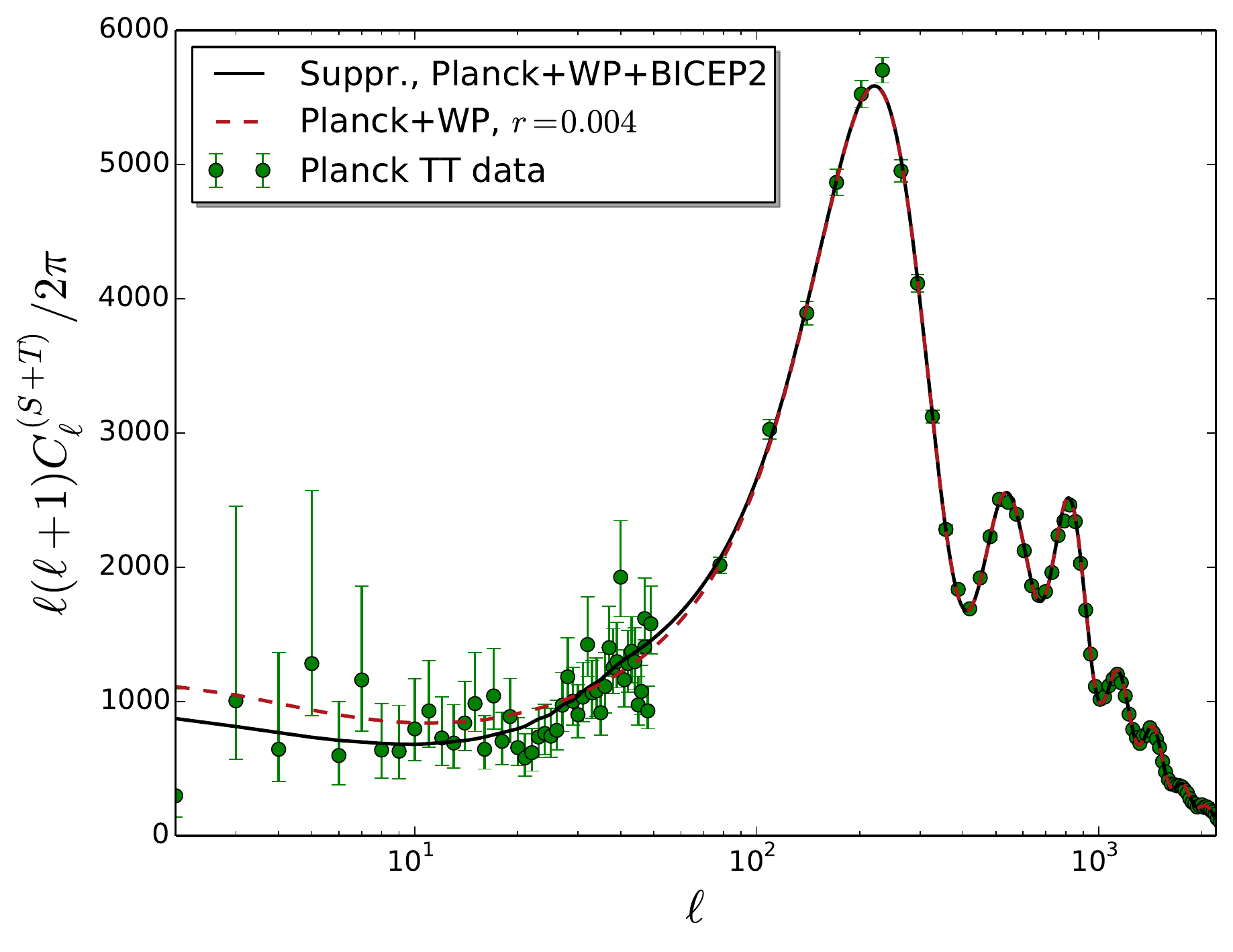}
\centering
\caption{The CMB multipoles as shown in \cite{contaldi}. The upper  line is the $\Lambda$CDM model best fit with constant $\ns=0.96$.
 The lower gives the best fit of \cite{contaldi}.}
\dlabel{cmb}
\end{figure}

The form \eq{calpofx} for $\zeta$ can arise if it is generated by  a curvaton-type field $\chi$ (one not affecting inflation) whose contribution has a significant contribution
from super-horizon scales \cite{p13,ekc,andrewm}.
 Then
we have \cite{p13}
\be
\frac65 \fnl(k) \zetal(\bfx)=-\cz(k) +
\az(k) \hat\bfp \cdot \bfx/\xls +  \cdots
, \dlabel{zetal} \ee
where $\fnl(k)$ is the reduced bispectrum in the equilateral configuration and $\zetal$ is the super-horizon contribution to $\zeta$ evaluated to linear
order in $\delta\chi$.
To make $\fnl$ scale-dependent we need
  non-linear evolution of $\chi$ during inflation \cite{bntw}, corresponding to a
 potential which isn't quadratic.

The perturbation $\zetal$ generates contributions to the CMB quadrupole, which (barring a strong cancellation) must be $\lsim$ the observed quantity.
This places a lower bound on the magnitude of $\fnl(k)$. In the present  paper I revisit the bound, and its significance.

\section{Contributions to $\zeta$}

I first discuss the various contributions to $\zeta$, clearing up some minor
 discrepancies between \cite{ekc,p13,andrewm,misaocrit}, and a major discrepancy with the formula given in \cite{hassan0,hassan1,hassan2}.

If $\zeta$ is generated from the perturbation of a curvaton-type field $\chi$,
 the non-linear $\delta
N$ formula gives \cite{ss,lr}
\bea
\zeta(\bfx) &=&  N(\chi(\bfx)) - N(\chi_0) \\
&=& N'(\chi_0)) \( \dchis(\bfx) + \dchil(\bfx) \)
+ \frac12 N''(\chi_0)) \( \dchis(\bfx) + \dchil(\bfx) \)^2
+\cdots \\
&\equiv& \( \zetas(\bfx) + \zetal(\bfx) \)
 + \frac35\fnl(k)
\( \zetas(\bfx) + \zetal(\bfx) \)^2 + \cdots
 \\
&=& \zetas(\bfx) + \frac35\fnl(k)  \zetas^2(\bfx)
 + \frac65\fnl(k) \zetal(\bfx) \zetas(\bfx) \nonumber \\
&& +\zetal(\bfx) + \frac35 \fnl(k)  \zetal^2(\bfx) +\cdots
, \dlabel{final} \eea
where
$\zetas \equiv  N'(\chi_0) \dchis(\bfx)$ and
 $\zetal \equiv N'(\chi_0)\dchil(\bfx)$.  I have equated  $N''/N'^2$ with
 $(6/5)\fnl(k)$, where $\fnl(k)$ is the reduced bispectrum of $\zeta$ in the equilateral configuration, which will be justified shortly.

To explain the dipole anisotropy, this expression is taken  to apply within a comoving box a bit bigger than $\xls$.
The left hand side is time-independent, but every object on the right hand side depends on time, which is chosen to be the epoch $t_k$ of horizon exit
  $k=a(t_k) H(t_k)$ for a scale $k\gg \xls\mone$. I am using  $k$ instead of $t_k$ to denote the epoch.
The unperturbed quantity $\chi_0$ is the average  within the  box.
The perturbations $\dchis$ and $\dchil$ are evaluated on the flat slicing. The perturbation $\dchis$ is generated by the vacuum fluctuation
on scales $1/k'\leq 1/k$, and the perturbation $\dchil$ comes from scales $1/k' \gg \xls$ (super-horizon scales).

The first two terms of \eq{final} give $\zeta_\bfk$. The dominant first term gives the spectrum
$\calpz(k)=
N'^2(\chi_0(k)) (H(k)/2\pi)^2$.
Since $\calpz(k)$ and $H(k)$ are both slowly varying, so is $N'$.
Including the second term, one can verify \cite{bntw} that
 $\fnl(k)$ is indeed
 the reduced bispectrum in the equilateral configuration.\footnote
 {This is Eq.~(3.5) of \cite{bntw}, evaluated in the equilateral configuration with  $t_i$ the epoch of horizon exit and  with  the second term
 set to zero. The justification for ignoring the second term is given, with references, in the
 paragraph following the one containing
 Eq.~(3.5).}

Omitting the small middle term, the first line of \eq{final} gives $\zeta(\bfx,k)$, which means that
\be
\frac65 \fnl(k) \zetal(\bfx) =- \cz(k) + \az(k) \hat \bfp\cdot\bfx/\xls + \cdots
, \dlabel{zetaleq}  \ee
where the dots indicate  sub-dominant  quadratic  and higher terms.
 I have ignored the dependence of $\zetal$ upon $t_k$.
 This is justified  because  $\zetal$ must be very slowly varying, since it gives the dominant
 contribution to $\zeta$ on super-horizon scales which itself is independent of $t_k$.
We see that $\cz$ and $\az$ are both proportional to $\fnl$.

In \cite{hassan0,hassan1,hassan2} they give a formula (Eq.~(15) of \cite{hassan0}) which seems to be  equivalent to \eq{zetaleq}, with the crucial difference that
$\fnl(k00)$ there denotes the squeezed configuration.
But the formula is not  really equivalent, because their
               quantity $\calr(t,\bfx)$, defined in \cite{hassan0}, is not the same as $\zeta$. That is because their $\calr(t,\bfx)$  is defined in  Eq.~(2) which refers to the  slicing of uniform  {\em curvaton} field perturbation. In contrast, $\zeta(t,\bfx)$ is defined by that formula, but referring to the slicing of uniform energy density. During inflation, the curvaton field is supposed to have a negligible effect which means that the slicing of uniform energy density is
                 the same as the slicing of uniform {\em inflaton} field \cite{withmisao}.
                                  Even if we identify their $\calr(\bfx)$ with our $\zeta(\bfx)$ (where now the argument $t$ is omitted because we deal with the late-time quantity that is
                 constrained by observation)
                    Eq.~(9) of \cite{hassan0} is not valid. This equation is valid for $\zeta$ during inflation, as discussed in \cite{cz}. It is not valid for the observed
                    $\zeta$ when that is generated by the perturbation of a curvaton-type field. As they recognise, their Eq.~(9) is the key to the rest of their analysis, which is therefore invalid.

                    In this paper I find that the effect being considered generates the equilateral configuration, not the squeezed configuration. That is despite the fact that the  effect comes from a  field perturbation on a super-horizon scale.

Although it is possible for a curvaton-type model to give the desired result \eq{zetaleq}, one should be clear that it is not easy. Indeed, when the rate of change of $\fnl$ is small it is given by \cite{byrnesetal}
\be
\left| \frac{d\fnl(k)}{d\ln k} \right|= \frac56\sqrt\frac r8 \frac{\mpl|V'''|}{3H^2}
, \dlabel{fnlexp} \ee
where $r\simeq 0.16$ is the tensor fraction, $H\simeq 1\times 10^{14}\GeV$ is the Hubble parameter during
inflation, $\mpl=2\times10^{18}\GeV$ is the reduced Planck mass and $V(\chi)$ is the potential of the curvaton-type field. We need $|V'''|$ to small enough to keep the rate of change small  while $\zeta$ is being generated on scales
$k\mone\lsim \xls/60$, but we need $|V'''|$ to be much bigger before that.
 This difficulty is of course merely a reflection of the fact that we are talking about  features that is present only on scales $k\mone \gsim \xls/60$,
 and any other explanation of the features would have a similar difficulty.

\section{The forms of $\az(k)$ and $\cz(k)$}

In the next section I am going to produce lower bounds on $|\fnl(k)|$ in terms of $\az(k)$ and
$\cz(k)$. To obtain numerical values for the bound I will have to assume something about
the latter quantities, which are both proportional to $\fnl(k)$.

 Within the large observational uncertainty, the suppression of multipoles can be fitted with a step function \cite{contaldi},
\be
\calpz(k) = \calpz(k_0)\( 1 - K \theta(k_0-k) \)
, \dlabel{step} \ee
with a best fit corresponding to
 $K=0.35$ and  $k\mone_0 < \xls/38$.
 This would correspond to $\cz(k)=0.17\theta(k_0-k)$.
 Taken literally, the step function \eq{step} would correspond to a delta function for $V'''$
   in \eq{fnlexp}
   but a monotonic
 variation of $V'''$ may also be  compatible with the data.

 A step function for $\cz(k)$ implies a step function for $\az(k)$, and I will take
  $\az(k)=0.07\theta(k_0-k)$. To cover the full range $\ell<60$ on which the anisotropy exists we need
  $k_0\mone=\xls/60$, which is a bit smaller than the value in the previous paragraph.
In view of the large uncertainties in the data I will ignore the difference, and for definiteness adopt
$k_0\mone=\xls/60$.

Instead of the step function forms for $\az$ and $\cz$ that I am adopting,  \cite{john4,john5} takes
$\az\propto\cz\propto k^{-n}$. To get sufficient suppression on scales $k\mone< \xls/60$ one needs $n>0.56$, but this also causes strong scale dependence
on scales $k\mone>\xls/60$.  For instance, setting $\cz(k)=0.15$ at the scale $k\mone=\xls/30$ and adopting
the minimal scale dependence $n=0.56$ gives complete suppression
($\cz=1$) at the scale $k\mone= \xls$.
It is shown in \cite{john4,john5} it is shown that the strong $k$-dependence causes $\calpz(k)$ to be  different from the CMB anisotropy parameter $A(\ell)$.

Although the  power-law form for $\cz(k)$ is very different from the step function form,
 the forms of the suppression of the multipoles might not be so different, at least if
    $n$ is not too far above $0.56$. In that case the power law form for $\cz(k)$ could be as viable as the step
    function. Using it would make little difference to my conclusions, the main effect being the reduction
    of the large-scale $\az(k)$ below the value $0.07$ that I am adopting. For $n=0.56$ the reduction
    according to \cite{john5} is by roughly one half.

\section{The EKC and GZ effects}

The EKC and GZ effects refer to the quadrupole $a_{2m}$ of the CMB anisotropy.
The EKC effect \cite{ekc} comes from the last term of \eq{final}.
To bound it one uses the data analysis of
 \cite{george} which gives
\be
  \sqrt{ \frac15\sum|a_{2m}|^2} =6.5\times 10^{-6}
. \dlabel{cbound1} \ee
 Using the Sachs-Wolfe approximation
\be
\Delta T(\bfe)/T = (1/5) \zeta(\xls\bfe)
, \ee
and requiring that the EKC contribution to the left hand side of \eq{cbound1} is less than the total,
one finds \cite{misaocrit}\footnote
 {The normalization of this result is almost the same as the one given in \cite{ekc}, who integrate an approximate evolution equation.
  (Both their procedure and the Sachs-Wolfe approximation ignore the primordial anisotropic stress of the neutrinos which gives an error of tens of percent.)
     Both \cite{ekc} and \cite{misaocrit} take $\fnl$ to be a constant.
  It was pointed out in \cite{p13} that the result holds in any curvaton-type model (ie.\ a model that generates $\zeta$ from the perturbation of a field different from the inflaton), allowing $\fnl(k)$.
   The normalization in \cite{p13} is  a bit too low though,
  because it referred to the arXive v1 of that paper which was later corrected; I thank A.\ Erickcek for clarifying this.}
\be |\fnl(k)|  \gsim 66 \( \frac{\az(k)}{0.07} \)^2
\dlabel{ekcbound}.\ee

The GZ effect \cite{gz} is the contribution to the CMB quadrupoles
 $a_{2m}$ coming from the quadratic part of $\zeta_L$, through the next-to-last term of \eq{final}.\footnote
{Contrary to the statement in  \cite{misaocrit}, the linear part of the next-to-last term of \eq{final} has no physical effect. This is because
the  linear contribution to  $\zeta$ has no physical effect by virtue of the adiabatic initial condition, which is valid in our setup. That is
the case for both a flat \cite{mb} and an open \cite{zsb} universe. Contrary to the statement in \cite{p13}, the EKC effect has nothing to do with the
GZ effect.}
                    To estimate the GZ effect,  one has to assume that we occupy a typical location, within some box large enough to contain the wavelengths that contribute to $\dchil$.
                    For simplicity I will assume that there is just one wavenumber $\kl$ and write.
                    \be
\calpzl(k) = P^2 \delta(\ln k-\ln \kl )
\ee

Equating $\zeta\sub L^2(0)$ with $\vev{\zeta\sub L^2}$ we have
\be
|\cz(k)| = \frac 65 P|\fnl(k)|
\dlabel{one}.\ee
Equating  $|\nabla\zetal(0)|^2$ with
$\vev{|\nabla \zetal|^2}$
 we find
\be
P\frac{\kl}{\ah}
 =  \left| \frac{\az(k)}{3.7 \fnl(k) } \right|
,\dlabel{two} \ee
where I used
$1/H_0=4,500\Mpc$ instead of $\xls$ with an eye to the open universe calculation.
Since $\az(k)$ is proportional to $\fnl(k)$ the right hand side is independent of $k$.

Using \eqs{one}{two},
\be
\kl/\ah= 0.23 \left| \frac{\az(k)}{\cz(k)} \right|
\dlabel{three}\ee

We also need the $C_2\equiv \vev{|a_{2m}|^2}$.\footnote
{This is the standard definition of $C_2$ (see for instance \cite{george}. In \cite{ekc,misaocrit,andrewm}, $C_2$ is instead taken to
denote the observed quantity $\frac15\sum |a_{2m}|^2$.}
In a flat universe, the
GZ contribution to
\be
\cgz=0.21 \( \frac \kl\ah \)^4 P^2
\dlabel{cexpression} \ee
Let us assume that
\be
\sqrt{\cgz}\lsim
\sqrt{ \frac15\sum|a_{2m}|^2} =6.5\times 10^{-6}
. \dlabel{cbound}
\ee
 Then  \eqss{two}{three}{cexpression}  need\footnote
 {This may be compared with the bound obtained in \cite{p13}, where only $|\cz|\ll 1$ was assumed. Setting $\cz=1$ in my expression, it
 is three times bigger (within a rounding error) than the one found in \cite{p13}, because the latter
paper used a bound on $\sqrt{\cgz}$ that was a factor 3 looser than \eq{cbound}.}
\be
\left| \fnl(k) \right|  \gsim 170 \( \frac{0.17}{|\cz(k)|} \)   \( \frac{\az(k)}{0.07} \)^2
\dlabel{fnlflat}.\ee

The open universe case has been considered in
\cite{andrewm}, where they point out that the super-horizon contribution
is generated in a previously proposed
 open universe inflation model \cite{open}.
  This slow-roll model is incompatible with
 complete dominance of the curvaton-type contribution, but it could allow a significant
 curvaton-type contribution which would be enough for our purpose. Let us proceed without tying
 ourselves to any particular inflation mechanism.

For an open universe, it is convenient to choose $1-\Omega_0 = 1/(\ah)^2$.
 Then curvature scale corresponds to $k=1$.
The previous result applies if $\kl\gg 1$ (sub-curvature scale), but things are different
if  $\kl\ll 1$ (super-curvature scale).
 Instead of \eq{cexpression} we have
 \cite{ourgz}
\be
\cgz=0.00543 \(1-\Omega_0 \)^2 \kl^2 P^2
\dlabel{open1}, \ee
and \eq{cbound} gives
\be
(1-\Omega_0) \kl P < 8.8\times 10^{-5}
\dlabel{open2} \ee
and \eq{two} can be written
\be
\( 1-\Omega_0 \)\half \kl P =   \left| \frac{\az}{3.7 \fnl} \right|
.\ee

After using \eq{one} to eliminate $P$, \eqs{open1}{open2} become
\bea
\( 1-\Omega_0 \) \kl &<& \frac65 \frac{\left| \fnl \right| }{\left| \cz \right|}\times 8.8\times 10\mfive \\
\( 1-\Omega_0 \)\half \kl &=& \frac65 \( \frac{\left| \az \right| }{3.7 \left| \cz \right| } \) \dlabel{opensecond}
.\eea

Eliminating $\kl$ from these gives
\be
\frac56 \left| \fnl \right| > \(\frac{ \( 1- \Omega_0 \)\half }{8.8\times 10\mfive}\)
 \(\frac{\left| \az \right| }{3.7} \)
\ee
Using $\kl\ll 1$, \eq{opensecond}  becomes
\be
\( \frac65 \) \frac{ \left| \az\right|  }{3.7\left| \cz \right| } \frac1{\(1-\Omega_0 \)\half } \ll 1
, \ee
leading finally to
\be
\left| \fnl \right| \gg 48 \( \frac{0.17}{|\cz(k)|} \)   \( \frac{\az(k)}{0.07} \)^2
. \ee
This is a bit different from the result in \cite{andrewm} because they don't use \eq{one} and just require
$P<1$.

The tightest bound is for the flat case, corresponding to the GZ bound \eq{fnlflat}.
It can be reduced from the fiducial value 170 by a factor 2 if we reduce
 $\az(k)$ by $1$-$\sigma$, or if we take on board the reduction of $\az(k)$ found in \cite{john5}. It  can be further
 reduced by say a factor three if we multiply the right hand side
 of \eq{cbound} by a factor three as was done in \cite{p13}. (Such a factor allows for some degree of cancellation between the GZ effect and the ordinary contribution to the quadrupole, and for the fact that we might not occupy a typical position within the large box.)  That would give $|\fnl|<30$.

Current observational bounds on $\fnl$ take it to be independent of the overall scale. Depending on the shape
that is adopted, one finds \cite{planckng} $|\fnl|\lsim 10$ to 100. As the $\fnl$ that we consider
falls off rapidly as one goes below the scale $k\mone\sim \xls/60$,
a new analysis of the CMB data will be needed to see
 if it is allowed.  Ideally such an analysis should use the full $\fnl(k_1,k_2,k_3)$ provided by a specific
 curvaton-type model, but for an estimate it may be good enough to identify $\fnl(k)$ with a suitable
 average over the $k_i$.

\section*{Acknowledgement}
I thank Andrew Liddle and John McDonald for useful comments on a draft of this work.

\end{document}